\documentclass[flushrt,preprint,12pt]{aastex}
\tighten
\def\LCDM{\Lambda\rm{CDM}}

\begin{document}

\title{Strong Gravitational Lensing and Dark Energy} 
\author{Nick Sarbu\altaffilmark{1,2}, David Rusin\altaffilmark{1} 
and Chung-Pei Ma\altaffilmark{1,2}}

\affil{$^1$ Department of Physics and Astronomy, University of
Pennsylvania, Philadelphia, PA~19104} 

\affil{$^2$ Department of Astronomy, University of California,
Berkeley, CA 94720 \email{insarbu@hep.upenn.edu; drusin@hep.upenn.edu;
cpma@astron.berkeley.edu}}


\begin{abstract}
We investigate the statistics of gravitational lenses in flat,
low-density cosmological models with different cosmic equations of
state $\omega$.  We compute the lensing probabilities as a function of
image separation $\theta$ using a lens population described by the
mass function of Jenkins et al. and modeled as singular isothermal
spheres on galactic scales and as Navarro, Frenk \& White halos on
cluster scales.  It is found that COBE-normalized models with $\omega
> - 0.4$ produce too few arcsecond-scale lenses in comparison with the
JVAS/CLASS radio survey, a result that is consistent with other
observational constraints on $\omega$.  The wide-separation
($\theta\ga 4''$) lensing rate is a particularly sensitive probe of
both $\omega$ and the halo mass concentration.  The absence of these
systems in the current JVAS/CLASS data excludes highly concentrated
halos in $\omega\la -0.7$ models.  The constraints can be improved by
ongoing and future lensing surveys of $>10^5$ sources.
\end{abstract}


\section{Introduction} 

A number of observational data sets strongly suggest that the
cosmological energy density includes a component that is not
associated with matter, either baryonic or dark.  For example,
measurements of the acoustic peaks in the cosmic microwave background
(CMB) power spectrum point to a nearly spatially-flat cosmology (Pryke
et al.\ 2001; Netterfield et al.\ 2001 and references therein).
However, the matter density $\Omega_m$ inferred from cluster baryon
fractions (e.g., David, Jones \& Forman 1995) and galaxy redshift
surveys (Percival et al.\ 2001 and references therein) falls far short
of the critical value.  Type Ia supernovae studies offer independent
evidence of a negative-pressure component of the energy density (Riess
et al.\ 1998, 2001; Perlmutter et al.\ 1999).  This energy component
is described by the equation of state $p = \omega \rho$ with $\omega <
0$.  The value of $\omega$ is $-1$ for the spatially homogeneous
cosmological constant $\Lambda$, and $\omega$ can be larger than $-1$
for other types of fields such as the quintessence (e.g., Frieman et
al.\ 1995; Turner \& White\ 1997; Caldwell et al.\ 1998).  The latter
clusters spatially on large scales, thereby modifying both the matter
density fluctuation power spectrum and the CMB anisotropy (Ma et al.\
1999).  Large-scale structure and supernova observations currently
favor $-1 \le \omega \la -0.6$ (Perlmutter, Turner \& White 1999; Wang
et al. 2000).  The constraint can be improved with ongoing and future
surveys based on, for example, the classical method of measuring the
luminosity distance or the differential volume element as a function
of redshift (e.g., Newman \& Davis 2000).

In this {\it Letter} we focus on strong gravitational lensing and
examine the effect of the equation of state $\omega$ on the
probabilities of producing multiply lensed systems.  The number of
expected lenses as a function of image separation provides a potential
means of constraining $\omega$ because it is determined by factors
such as the angular diameter distances to the lens and the source, the
lensing cross sections, and the number density of the lenses, each of
which depends on $\omega$.  This type of study is timely in view of
the completion of the Jodrell-VLA Astrometric Survey (JVAS; e.g.,
Patnaik et al.\ 1992; King et al.\ 1999) and the Cosmic Lens All-Sky
Survey (CLASS; e.g., Myers et al.\ 1999, 2001), which together provide
the largest uniformly selected sample of radio lens systems and have
yielded 18 new lenses thus far.  The Sloan Digital Sky Survey (SDSS)
will further increase the source population by one to two orders of
magnitude.

Since lensing probes the mass and not the light distribution, we model
the lenses as a population of dark matter halos with an improved
version of the Press-Schechter (1974) mass function.  For galaxy-size
lenses, we follow the tradition in lensing studies and use the
singular isothermal spheres (SIS) as the mass profile (e.g., Turner et
al. 1984; Narayan \& White 1988; Kochanek 1996).  The SIS ensures flat
rotation curves and is consistent with constraints on the inner mass
profiles of elliptical galaxies (e.g., Rusin \& Ma 2001; Rix et
al. 1997; Kochanek 1995; Cohn et al. 2001).  For cluster-size lenses,
however, the mass profile is mostly determined by the dark matter.
For this, we take advantage of the recent progress in high resolution
$N$-body simulations and model the lenses with the phenomenological
profile of Navarro et al. (NFW, 1997).  This approach allows us to
calculate both small-separation (galaxy-based) and wide-separation
(dark matter-based) lensing phenomena concurrently.  It also enables
us to relate the lensing probabilities directly to cosmological
parameters through the mass power spectrum that governs the number
density of lenses, and the lensing cross sections.  Approximating
lenses with a mixture of SIS and dark matter profiles is supported by
simple baryon cooling models (Keeton 1998; Kochanek \& White 2001) and
has been used recently to study the link between strong lensing and
distant supernovae (Porciani \& Madau 2000), halo density profiles
(Keeton \& Madau 2001; Takahashi \& Chiba 2001; Wyithe et al. 2001),
and $\Omega_m$ and $\Omega_\Lambda$ (Li \& Ostriker 2001).  These
studies, however, have assumed $\omega=-1$.  As we will show, relaxing
the $\omega=-1$ assumption can have a significant effect on the
lensing probabilities, in particular at wide angular separations.

\section{Lensing Probabilities and JVAS/CLASS Data}

In this {\it Letter} we consider cosmological models with a
present-day density parameter of $\Omega_m=0.35$ in matter (with 0.05
in baryons and 0.3 in cold dark matter (CDM)), $\Omega_{\omega}=0.65$
in dark energy, and Hubble parameter $h=0.65$.  For the dark energy
component, we analyze the standard cosmological constant ($w=-1$) and
models with $w< -1/3$.  This set of parameters is chosen to lie well
inside the concordance region given by large scale structure
observations (see Wang et al. 2000).  Taking into account the current
observational uncertainties in $\Omega_m$ and $h$ changes our
constraints on $\omega$ by up to $\sim$ 10\% (see \S~3).

The probability for lensing with an image separation greater than
$\theta$ is given by (e.g., Turner et al. 1984; Narayan \& White 1988)
\begin{equation}
  P ( >\theta )=\int dz_s {\cal P} (z_s)\int_0^{z_s} dz_l
    \frac{dr}{dz_l} \int^{\infty}_{M_{min}} dM\,
    \frac{dn}{dM}(M,z_l)\, \sigma_{\rm lens}(M,z_l)\,  B(M,z_l,z_s) \,,
\end{equation}
where $n(M,z_l)$ is the physical number density of dark halos at the
lens redshift $z_l$, $\sigma_{\rm lens}(M,z_l)$ is the lensing cross
section of a halo of mass $M$ at $z_l$, $B$ is the magnification bias
(see below), and $r$ is the proper cosmological distance to the lens,
with $dr/dz=cH_0^{-1}(1+z)^{-1}[\Omega_m (1+z)^3+ \Omega_\omega
(1+z)^{3(1+\omega)}]^{-1/2}$ for $\Omega_m+\Omega_\omega=1$.  The
parameter $M_{min}$ is the mass needed for a halo at $z_l$ to create
an angular separation $\theta$ between the outermost images of a
source at $z_s$.  By integrating over $z_l$ from $0$ to $z_s$, we
obtain the probability of lensing a source at $z_s$. The last integral
is over the redshift distribution ${\cal P}(z_s)$ of the sources (see
below).

For the halo number density, we use the recently calibrated expression
$dn/d\ln M=0.315\,(1+z)^3(\bar\rho/M) (d\ln\delta^{-1}_{\rm rms}/d\ln
M) \exp (-|\ln \delta_{\rm rms}^{-1}+0.61|^{3.8})$ by Jenkins et
al. (2001), where $\bar\rho$ is the mean mass density of the universe
and $\delta_{\rm rms}$ is the rms mass density fluctuation.  This
formula gives an improved match to the halo abundance found in
numerical simulations compared with the classic Press \& Schechter
(1974) formula, which tends to overestimate the abundance of typical
$M_*$ halos and underestimate massive halos.  In the calculation of
$\delta_{\rm rms}$, we use the COBE normalized mass power spectrum of
Ma et al. (1999) for the $w>-1$ models.  The mass is taken to be the
virial mass $M =4\pi\Delta_{vir}\bar\rho R^3_{vir}/3$, where $R_{vir}$
is the virial radius within which the average density is $\Delta_{vir}
\bar\rho$.  For the standard CDM model with $\Omega_m=1$,
$\Delta_{vir}$ is given by the familiar value $\Delta_{vir} =
18\pi^2\approx 178$; for flat $\LCDM$ models, $\Delta_{vir}$ can be
approximated by $\Delta_{vir} \approx (18\pi^2+82x-39x^2)/\Omega_m(z)$
with $x=\Omega_m(z)-1$ (Bryan \& Norman 1998).  Jenkins et al. has
specifically stated that their formula gives better fits to $dn/dM$
with $\Delta_{vir}=178$ regardless of the cosmological model, so we
followed this instruction.

The lensing cross section $\sigma_{\rm lens}$ in eq.~(1) depends on
the mass profile of the lenses.  Since our interest is to predict the
lensing probability over a wide range of image separations, we
consider both galaxy-size halos that are responsible for lens systems
of a few arcseconds, and cluster-size halos for larger $\theta$.  In
the former case, we approximate the galaxy mass distribution as an SIS
with a density profile $\rho(r)=\sigma_v^2/(2\pi G r^2)$, where
$\sigma_v$ is the 1-d velocity dispersion.  An SIS lens produces an
image separation of $2\theta_E$, where $\theta_E=4\pi (\sigma_v/c)^2
D_{ls}/D_s$ is the Einstein radius, and the cross section is
$\sigma_{\rm lens}=\pi (\theta_E D_l)^2 = 16\pi^3 (\sigma_v/c)^4
(D_lD_{ls}/D_s)^2$ (e.g., Schneider et al. 1992), where $D_s, D_l$,
and $D_{ls}$ are the angular diameter distances to the source, to the
lens, and between the lens and the source, respectively.  Using
$\sigma_v=(\pi G^3 M^2 \Delta_{vir}\bar\rho/6)^{1/6}$, we can then
relate $\sigma_{\rm lens}$ to the halo mass which is needed for
eq.~(1).

For lensing by cluster-size halos, we approximate the cluster mass
distribution by the density profile determined from halos in numerical
simulations (Navarro et al. 1997): $\rho(r)=\bar\rho\,\bar\delta\,
u(r/r_s)$, where $u(x)=1/[x(1+x)^2]$.  This profile is shallower than
the SIS for the inner parts of a virialized halo but is steeper at
large radii.  Other than the virial mass, this profile is described by
a concentration parameter $c \equiv R_{vir}/r_s$, where $R_{vir}$ is
the virial radius of the halo discussed earlier, and $r_s$ is a scale
radius.  For $c$, we take into account both the mass and redshift
dependence and use the relation of Bullock et al. (2001): $c(M,z)=9
(1+z)^{-1} (M/M_*)^{-0.13}$, where $M_*=1.5\times 10^{13}
h^{-1}M_\odot$.  The density amplitude $\bar\delta$ is related to $c$
by $\bar\delta=\Delta_{vir} c^3 [\ln(1+c)-c/(1+c)]^{-1}/3$.  The
lensing cross section for NFW halos is determined by the parameter
$\kappa_0=\bar\rho \bar\delta\,r_s/\Sigma_{cr}$, where
$\Sigma_{cr}=(c^2/4\pi G)(D_s/D_l D_{ls}) $ is the critical surface
mass density.  An NFW halo will have multiple images if the source is
within the radial caustic of angular size $\beta_{rad}$ from the halo
center.  We compute $\beta_{rad}$ as a function of halo mass, lens
redshift, and source redshift by solving $d\beta/d\theta=0$, where
$\beta$ and $\theta$ are the positions of the source and the image,
respectively.  The cross section for lensing by an NFW halo is then
$\sigma_{\rm lens}=\pi (\beta_{rad} D_l)^2$.  We also need to
determine the halo mass $M_{min}$ in eq.~(1).  Here we use the fact
that the angular separation of the outermost images is insensitive to
the value of $\beta$ (Schneider et al. 1992) and simplify the
calculations by using $\beta=0$ for a perfectly aligned source-lens
configuration.  We assume a ``cooling mass'' of $M_c \approx 1.5\cdot
10^{13} h^{-1} M_{\odot}$, above which the lenses are assigned the NFW
profile and below which the lenses are SIS.  Uncertainties introduced
by $M_c$ on the lensing probability are discussed in \S~3.

To compare the predicted lensing probabilities with observational
results, we use the combined data from JVAS and CLASS, which offer the
largest uniformly selected sample of gravitational lens systems. These
surveys have discovered 18 lenses among a sample of $\simeq$ 12000
flat-spectrum radio sources (Browne \& Myers 2000; Browne et al.\
2001; Myers et al.\ 2001). A robust statistical analysis requires that
careful cuts be made to the above sample, and this will be discussed
in detail by Browne et al. However, because preliminary estimates
indicate that the lensing rate will not differ much from the $\simeq
1/600$ value derived here, we will assume a sample of 18 lenses and
12000 sources in the present analysis.

Raw optical depths must be corrected to account for the magnification
bias, which leads to an over-representation of lensed sources in any
flux-limited sample (e.g., Turner et al.\ 1984; Maoz \& Rix
1993). Magnification bias enhances the lensing probability of sources
in a bin of total flux density ($S$) by the factor $B(S) =
\phi^{-1}(S) \int d\mu\,\mu^{-1} P(\mu) \phi(S/\mu)$, where $\phi(S)$
is the source luminosity function and $P(\mu)$ is the distribution of
total magnifications ($\mu = \sum_i |\mu_i|$, where the magnification
of the $i$th image is $\mu_i$) produced by the lens.  The sources
probed by CLASS are well-represented by a power-law luminosity
function, $\phi(S) = dn/dS \propto S^{-\eta}$, with $\eta \simeq 2.1$
(Rusin \& Tegmark 2001).  The bias thus reduces to a simple form that
is independent of flux density: $B = <\mu^{\eta-1}>$.  An SIS lens
produces total magnifications described by the probability
distribution $P(\mu) = 8 \mu^{-3}$; so if $\eta = 2.1$, $B_{SIS} =
4.76$.  The cross sections of all SIS lenses are enhanced by this
factor.  The situation is more complicated for the NFW profile as its
lensing properties depend on $\kappa_0$, which in turn depends on the
halo mass and the angular distances to the halo and the source.  We
compute numerically the magnifications produced by NFW halos for $0.1
< \kappa_0 < 10$ and use this to tabulate $B_{NFW}(\kappa_0)$.

For the sources in the JVAS/CLASS survey, the redshift distribution is
still poorly understood but the mean redshift is estimated to be
$<z_s>=1.27$ (Marlow et al. 2000).  Since the lower flux source
distribution of the JVAS/CLASS survey is indistinguishable from the
complete, brighter source distribution of the Caltech-Jodrell Bank
VLBI sample (Henstock et al. 1997), we assume the latter quasar
distribution for ${\cal P}(z_s)$ in eq.~(1).

\section{Results and Discussion}

Fig.~1 compares the JVAS/CLASS data with the predicted lensing
probabilities $P(>\theta)$ calculated from eq.~(1) for the $\LCDM$
model and models with $\omega=-2/3, -1/2, -1/3$.  The probability at
$\theta\ga 4''$ decreases rapidly as $\omega$ increases towards 0.
The most important systematics that affect wide separation lensing are
due to the halo concentration parameter $c(M,z)$. In Fig.~2a we
quantify the dependence of $P(>4'')$ on both $\omega$ and the
coefficient of $c(M,z)$, $c_* \equiv c(M_*,z=0)$.  It shows that
models with larger $\omega$ can tolerate a higher halo concentration
due to the lower lensing rates in these models. JVAS/CLASS thus far
has detected no lens systems with $\theta >4''$ (Phillips et al.\
2000). This excludes the highly concentrated halos in $\omega\la -0.7$
models as indicated by the $1\sigma$, 90\%, and 95\% confidence
contours in Fig.~2a. We also plot the $1\sigma$ confidence levels that
would be imposed if one wide separation lens system were discovered.
We interpret the result as indicating that higher large separation
lensing rates would lead to more refined constraints in the
$\omega-c(M)$ plane.

We can also compare the total number of lenses predicted by the models
with the 18 systems found among $\sim 12000$ sources in JVAS/CLASS.
We find the expected number of lenses with $\theta > 0.3''$ (the
approximate angular resolution of JVAS/CLASS) to be $20.7$, $18.2$,
$15.5$ and $9.6$ for the $\LCDM$, $\omega=-2/3$, $-1/2$, and $-1/3$
models shown in Fig.~1, respectively.  Because the SIS lensing cross
section is several orders of magnitude higher than that of the NFW,
the expected number of lenses is not sensitive to the concentration
$c(M)$.  Instead, it depends more strongly on the cooling mass $M_c$
because a larger $M_c$ allows more halos to be modeled as SIS.  In
Fig.~2b we quantify the dependence of the total lensing rate on $M_c$
and $\omega$ by plotting the relative likelihood curves
$L(\omega)\propto p(\omega)^{N_l} (1-p(\omega))^{N-N_l}$ for detecting
$N_l=18$ lenses from $N=12000$ sources, where $p(\omega)$ is the
model-predicted lensing rate for a given $\omega$.  The range of $M_c$
is similar to that discussed in Kochanek \& White (2001), with the
upper and lower limits close to the cooling masses used by Porciani \&
Madau (2000) and Li \& Ostriker (2001).  The $2\sigma$ confidence
level in Fig.~2b suggests that the constraint on the equation of state
is not sensitive to moderate alterations to the cooling mass.  Models
with $\omega \ga -0.4$ are disfavored in all three cases.

Note that our calculations implicitly assume that each halo below
$M_c$ harbors sufficient baryons to give an SIS profile for the {\it
total} mass.  It is likely, however, that lower mass halos be devoid
of baryons and therefore retain a shallower profile (e.g., NFW).
Because an SIS lens has a much larger cross section than an NFW lens
of the same mass, modeling all low-mass halos as SIS would
overestimate the total lensing rate.  Constraints on $\omega$ derived
under our assumptions are therefore conservative -- if low-mass halos
actually contribute smaller cross sections than our calculations
predict, more negative $\omega$ would be needed to reproduce the
observed lensing rate.

Both large separation lensing probabilities and the total predicted
number of lenses are affected by the cosmological parameters $h$ and
$\Omega_m$. Our investigations show that varying these parameters
within the $1\sigma$ error limits given in Netterfield et al. (2001)
leads to almost parallel shifts in the contour lines and likelihood
curves in Fig.~2 by $0.075$ or less in $\omega$, therefore not
affecting the generality of our results.

As the figures show, the predicted lensing probability decreases as
$w$ changes from $-1$ towards 0.  The effect is particularly strong
for lenses with $\theta\ga 4''$, where the probability decreases by
more than two orders of magnitude when $\omega$ is varied from $-1$ to
$-1/3$.  The dependence on $\omega$ enters in two ways in our
calculations: one through the mass power spectrum upon which the halo
density $dn/dM$ depends; the other through kinematic factors
such as the angular diameter distances which are functions of
$\omega$.  For the power spectrum, different $\omega$ gives different
linear growth rate for the density field.  Specifically, for a fixed
$\Omega_m$ and $\Omega_w$, the growth becomes slower as $\omega$
increases towards zero because the energy density in quintessence-type
of fields dominates over that in matter at earlier times, resulting in
earlier cessation of the gravitational collapse (Ma et al. 1999).  The
amplitude of the COBE-normalized power spectrum is generally lower for
larger $\omega$, resulting in a smaller halo density and hence a lower
lensing probability.  Besides the power spectrum, the dependence of
the lensing probabilities on $\omega$ also enters through the
effective lensing volume element $(D_l D_{ls}/D_s)^2 (dr/dz)$, where the
first factor comes from the lensing cross section $\sigma_{\rm lens}$
and the second factor is the distance to the lens
in eq.~(1).  These effects are illustrated in Fig.~3.

It is interesting to compare gravitational lensing studied in this
{\it Letter} with classical cosmological tests for constraining the
equation of state.  Ongoing efforts such as high redshift supernova
searches and deep galaxy surveys offer promising ways to constrain
$\omega$ by determining the luminosity distance $d_L$ and the
cosmological volume element $dV/dzd\Omega$ (e.g., Newman \& Davis
2000; Turner 2001).  In Fig.~3 we compare the luminosity distance and
the volume element with the probability $P(>4'')$ for large separation
lensing.  As we can see, $P(>4'')$ depends more strongly on $\omega$
than the other quantities.  Gravitational lensing statistics from
ongoing and future surveys with $10^5-10^6$ sources (e.g., the SDSS;
see Cooray \& Huterer 1999) may offer an independent probe of
$\omega$, provided that systematics associated with input source
redshift distributions and halo concentration parameters can be
minimized.

This paper would not have been possible without the hard work of
everyone in the JVAS/CLASS team.  We thank Mike Turner for useful
comments.  C.-P. M. acknowledges support of an Alfred P. Sloan
Foundation Fellowship, a Cottrell Scholars Award from the Research
Corporation, a Penn Research Foundation Award, and NSF grant AST
9973461.

\clearpage
\begin{figure*}
\begin{center}
\begin{tabular}{c}
\epsscale{.8}
\plotone{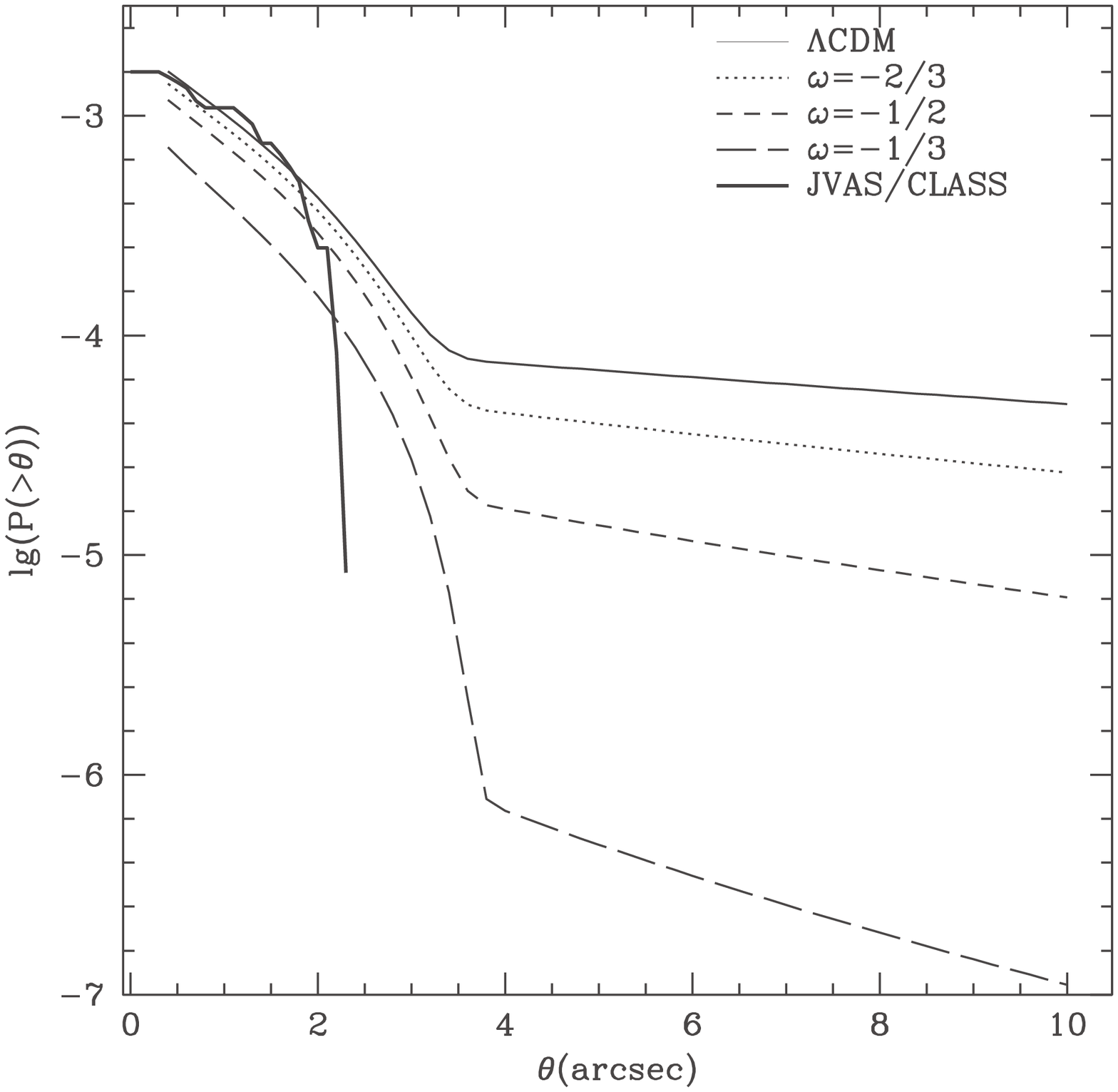}
\end{tabular}
\figurenum{1}\label{f1}
\caption{Predicted lensing probability with image separation $>\theta$
in cosmological models with different equation of state $\omega$.  The
lenses follow the mass function of Jenkins et al. (2001) and are
modeled as SIS for $M < 1.5\times 10^{13} h^{-1} M_\odot$ and as NFW
halos with the concentration parameter
$c(M,z)=9(1+z)^{-1}(M/M_*)^{-0.13}$, where $M_*=1.5\times 10^{13}
h^{-1} M_\odot$ (Bullock et al. 2001).  The histogram represents the
survey results of JVAS/CLASS.}
\end{center}
\end{figure*}

\clearpage
\begin{figure*}
\begin{center}
\begin{tabular}{c}
\epsscale{.65}
\plotone{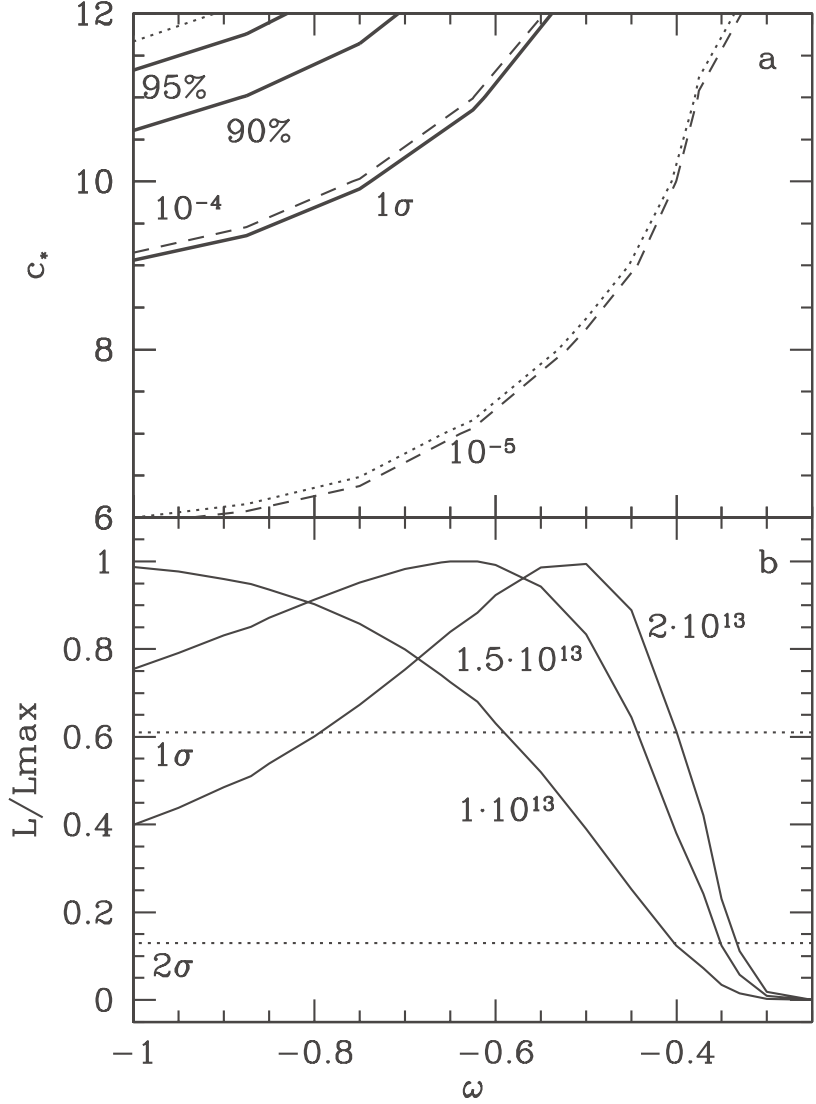}
\end{tabular}
\figurenum{2}\label{f2}
\caption{{\bf a)} Wide separation lensing probability $P(>4'')$ as a
function of the equation of state $\omega$ and the halo concentration
$c_*$, where $c(M,z)=c_* (1+z)^{-1}(M/M_*)^{-0.13}$.  Thick contours
are various confidence upper limits for JVAS/CLASS, which has detected
zero lenses with $\theta >4''$. (Upper-left corner is excluded.)
Dotted contours represent the $1\sigma$ confidence levels that would
be imposed if one large separation lens were discovered.  Dashed
contours are for $P(>4'')=10^{-4}$ and $10^{-5}$.  {\bf b)} Likelihood
curves for detecting 18 lenses from 12000 sources (found in
JVAS/CLASS) for three different cooling masses.  The
dotted lines represent the $1\sigma$ and $2\sigma$ confidence levels.
Models with $\omega \ga -0.4$ are disfavored at $2\sigma$.
}
\end{center}
\end{figure*}

\clearpage
\begin{figure*}
\begin{center}
\begin{tabular}{c}
\epsscale{.8}
\plotone{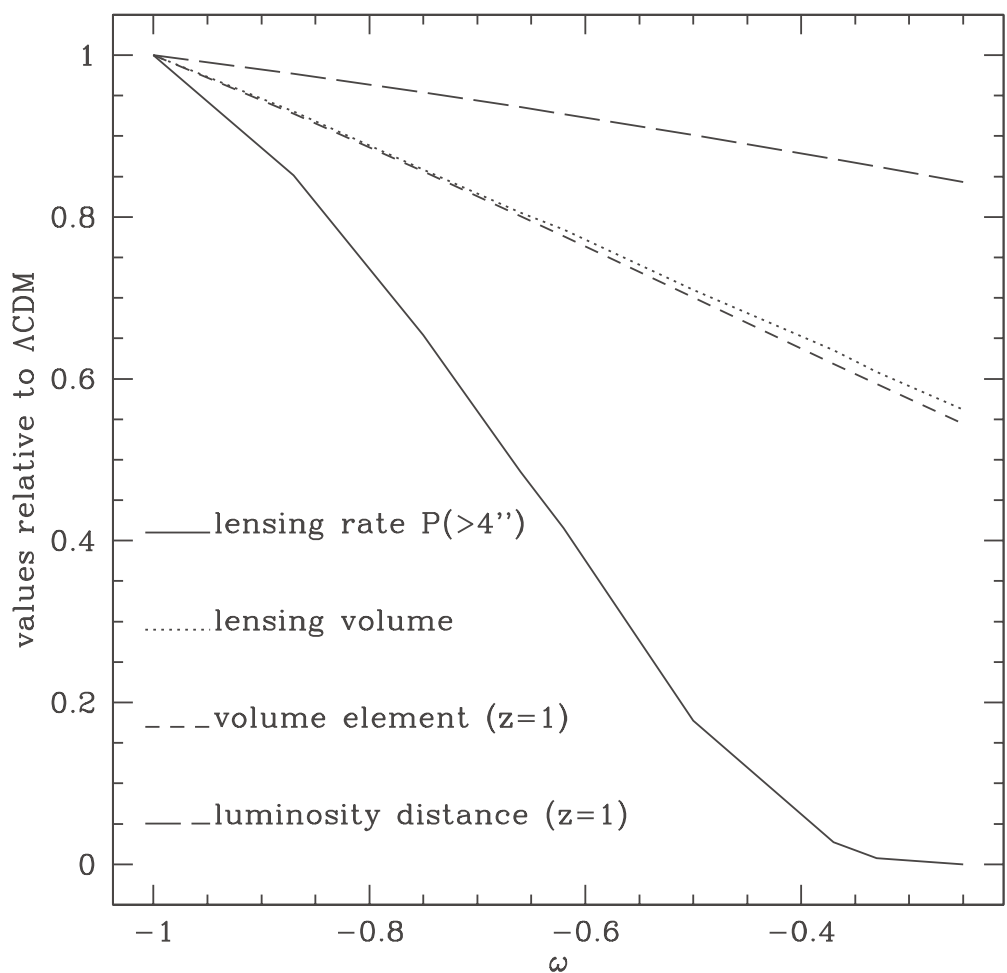}
\end{tabular}
\figurenum{3}\label{f3}
\caption{Ratio of four quantities in an $\omega>-1$ model relative to
the $\LCDM$ model: the lensing probability $P(>4'')$ for wide
separation systems, the effective lensing volume element $(D_l
D_{ls}/D_s)^2\, dr/dz$ (at $z_l=0.5$ and $z_s=1.5$), the volume
element $dV/dzd\Omega$, and the luminosity distance $d_L$ (both at
$z=1$).}
\end{center}
\end{figure*}

\end{document}